# Adaptive Speckle Imaging Interferometry: a new technique for the analysis of micro-structure dynamics, drying processes and coating formation.

L. Brunel[1*], A. Brun[1], P. Snabre[2] , L. Cipelletti[3]

*(1) FORMULACTION, 10 Impasse Borde Basse, F-31240 L'Union , France.*
*http://www.formulaction.com*

*(2) CRPP, CNRS, avenue Albert Schweitzer, F-33600 Pessac, France*

*(3) LCVN and Institut Universitaire de France, CNRS UMR 5587, Université Montpellier II, Pl. E. Bataillon 34095 Montpellier Cedex 05, France*

*\*Corresponding author: brunel@formulaction.com*

**Abstract:** We describe an extension of multi-speckle diffusing wave spectroscopy adapted to follow the non-stationary microscopic dynamics in drying films and coatings in a very reactive way and with a high dynamic range. We call this technique "Adaptive Speckle Imaging Interferometry". We introduce an efficient tool, the inter-image distance, to evaluate the speckle dynamics, and the concept of "speckle rate" (SR, in Hz) to quantify this dynamics. The adaptive algorithm plots a simple kinetics, the time evolution of the SR, providing a non-invasive characterization of drying phenomena. A new commercial instrument, called HORUS®, based on ASII and specialized in the analysis of film formation and drying processes is presented.

## 1. Introduction

Besides direct observation using a microscope, the first known optical technique to analyze the dynamical behavior of micrometric particle suspensions is Dynamic Light Scattering (DLS) [1,2], sometime also referred to as Photon Correlation Spectroscopy (PCS). The optical set-up consists in a point-like detector, usually a phototube or an avalanche photodiode, which collects the light scattered by the particles at a fixed angle with respect to the propagation direction of the incident light, usually a coherent laser beam. As the particles move (e.g. because of Brownian motion), the relative phase of the scattered photons change, resulting in temporal fluctuations of the detected intensity. These fluctuations are characterized by measuring their time auto-correlation function $g_2(\tau)=<I(t)I(t+\tau)>_{time}/<I^2>_{time}$, with $I(t)$ the time-dependent scattered intensity and $\tau$ the time delay. For Brownian particles under single scattering conditions, $g_2$ decays exponentially with a decay time that depends on the particles' diffusing coefficient [2], thus providing a popular particle sizing method. In 1987, Maret and Wolf [3] proposed an analytic formula for $g_2$ for Brownian scatterers in the opposite limit of strong multiple scattering, where photons undergo a large number of scattering events along their path through the sample. In 1988, Pine et al. [4] introduced the expression "Diffusing Wave Spectroscopy" (DWS) to designate DLS under strong multiple scattering conditions; since then, DWS has become a powerful and widespread technique for characterizing colloidal suspensions [5].

In traditional DLS and DWS techniques, achieving a good signal/noise ratio requires $g_2$ to be averaged over at least 1000 times its characteristic decay time. For samples with a slow dynamics (e.g. concentrated or aggregating suspensions) this may require a too long measurement time. Moreover, for non-stationary dynamics (e.g. in aggregation or drying processes) extensive time averaging may lead to incorrect results, if the dynamics evolves significantly during the measurement time. Recently, these limitations have been circumvented by sampling a large number of speckles. To fix the ideas, let us consider the case of a DWS experiment with a single point-like detector. As recalled above, about 1000 statistically independent samples are needed to build a good correlation function. These samples are usually collected sequentially in time, assuming that time averaging is equivalent to averaging in the ensemble of photon paths. This condition is normally met, since the scatterers move substantially and explore a significant fraction of all possible configurations during the measurement time. Thus time and ensemble averages are equivalent, a condition referred to as ergodicity. In other words, if the sample micro-structure evolves rapidly enough, a sufficiently large fraction of all possible light paths may be sampled in a reasonable time and time averaging yields the desired ensemble average.

For slow or non-stationary dynamics, various schemes have been proposed to speed-up this sampling process, in addition to introducing methods to correct for insufficient time-averaging [6]. In the cell-scanning technique [7] and the two cells technique [8] (or its rotating-frosty-glass variant [9]) a point-like detector is illuminated sequentially by different speckles, either by scanning the beam through the sample or by "scrambling" the scattered photons by letting them go through an additional scatterer before being detected (the 2$^{nd}$ cell or the rotating frosty glass). In the so-called "multispeckle" methods [9,14-17], by contrast, different speckles are sampled in parallel by using a large number of independent detectors, usually the pixels of a Charge Coupled Device (CCD) or CMOS camera (it is also possible to use a series of single photo-diodes or optical fibers coupled to photo-diodes). A correlation function is computed for each pixel, and the desired ensemble-averaged $g_2$ is obtained by averaging over all pixels: $g_2(\tau)=<I(t)I(t+\tau)>_{pixels}/<I^2>_{pixels}$. By adjusting the detection optics in such a way that the speckle size matches the pixel size (typically of the order of 10 µm), one ensures that each pixel carries a statistically independent contribution to $g_2$ and that the overall signal/noise ratio is maximized [9]. The acquisition time required to achieve a given precision, compared to that with a single detector, is reduced by a factor equal to the number of pixels. Since the latter is typically as large as several tens of thousands, in practice no time averaging at all is required, thus providing access to the instantaneous dynamics. Together with the capability of measuring slow and non-stationary dynamics, advantages include a simpler set-up and a cost per detector (pixel) very low. The limitations are the reduced speed and sensitivity of CCD and CMOS cameras compared to avalanche photo-diodes or phototubes. The maximum acquisition rate is 30Hz for a classical cheap camera and about a few kHz for more sophisticated and expensive cameras, limiting the smallest accessible delay to a fraction of msec at best, instead of about 10 nsec for traditional point-like detectors.

In this work we present three new developments of the multi-speckle technique. First, we propose a new tool to quantify the evolution of the speckle pattern recorded by a CCD or CMOS camera, which we term the inter-image distance. Second, we show that for most cases of practical importance in film formation and drying processes, the inter-image distance grows with time and saturates according to a simple exponential law. This allows us to introduce an efficient and quick way to quantify the rate of change of the speckle pattern which we term Speckle Rate, (SR). Third, we present an adaptive algorithm that continuously adjusts the camera acquisition rate to trace in real time the speckle dynamics with high reactivity, dynamic range and accuracy. We globally term "Adaptive Speckle Imaging Interferometry" (ASII) these three advances. ASII is thus an extension of multi-speckle (MS) DWS adapted to characterize film formation and drying processes by plotting a simple kinetics, the time evolution of the SR, providing a new, non-invasive signature of micro-structure evolution.

As a final introductory remark, we note that the temporal evolution of the speckle pattern generated by a sample with internal dynamics has also been investigated as a natural extension of the early works on the statistics of frozen speckles, often in connection with metrology and holography studies [10]. The scientific community involved in these works (oddly somehow separated from the DLS and DWS ones) refers to this phenomenon as to "dynamic speckle". The most popular way to characterize the "dynamic speckle" phenomenon is to measure the contrast $<I^2>/<I>^2$ of the speckle pattern, the underlying idea being that fast-moving speckles would lead to a blurred speckle pattern and hence to a reduced contrast. Both point-like [10] and CCD detectors [11] has been used in measurements over a variety of samples, ranging from biological tissues [10] to paint films [11] similar to those explored in our work. Compared to our approach, speckle contrast techniques yield less information, since the dynamics is probed on a single time scale, the integration time over which the detector is exposed to the speckle pattern. By contrast, in our technique the inter-image distance is calculated for time lags $\tau$ spanning several decades. As we shall show in the following, this is a crucial requirement for characterizing the dynamics of film, since the speckle rate may vary over 4 orders of magnitude.

More sophisticated approaches, involving the calculation of correlation functions similar to $g_2$ or to the inter-image distance introduced here have been used to characterize the drying of coatings in Refs. [12] and [13], respectively. In [12], a point-like detector was used, thus requiring a temporal average of the correlation function. This limits the effectiveness of this technique when the dynamics are very slow and non-stationary, as discussed above for DLS and DWS. In [13], a CCD detector was used. The second order moment, *IM*, of the so-called modified co-occurrence matrix was calculated, a quantity similar to the inter-image distance defined here. However, *IM* was calculated for a single fixed time delay, whereas in our work the inter-image distance is calculated for several $\tau$'s and the minimum lag is continuously optimized by varying the CCD acquisition rate. This allows us to characterize the film dynamics with great accuracy and in real time over more than four orders of magnitude in time, an achievement not possible with the method proposed in [13].

This short overview of previous works in the field of DLS and DWS and in that, closely related, of dynamic speckle interferometry shows that the ASII method proposed here represents a significant advance with respect to the existing optical techniques for the characterization of film formation and drying. The ASII method is at the core of a new commercially available instrument called HORUS®, which monitors the drying process of any product deposited on any substrate by analyzing the micro-structure dynamics of the film [18]. HORUS® provides the ink and paint industries with a new valuable tool for the non-intrusive monitoring of the drying and film formation process.

The paper is organized as follows: the hardware configuration is described in Section 2. Section 3 explains the ASII method: the inter-image distance, the exponential approximation and the adaptive algorithm are described here. Some examples of the application of HORUS® to the analysis of film formation are given in Section 4, before the concluding remarks of Section 5.

**2. Horus® hardware configuration**

A backscattering configuration has been chosen for the HORUS® apparatus, allowing the detector and the laser source to be housed in a single convenient optical head, as shown in Fig. 1. Typically, the backscattered signal is higher than the transmitted one, allowing one to use a detector with a relatively low sensitivity (such as a CCD or a CMOS camera) and limiting the required laser power. This last point is important for safety reasons and for preventing heat-induced convection in the sample. Indeed, in the backscattering geometry even a low-power laser diode allows one to monitor the drying of relatively black products such as black paints or inks. The product to be studied is spread out in a thin layer on any chosen substrate. Mechanical stability is important, since any vibration of the optical head would result in a spurious loss of correlation, wrongly attributed to the micro-structure dynamics of the analyzed product. The substrate is thus fixed on a massive support plate, which is carefully leveled with respect to gravity in order to avoid any flow. The optical head is attached to a rigid vertical mast fixed to the plate. The apparatus is designed so as to filter out any external vibration, limiting the maximum displacement of the head with respect to the plate to less than 1 μm. The head contains a CMOS camera detector (with no lens) and the laser source. A hole acting as a diaphragm blocks the parasitic ambient light coming from outside a cone of half-angle approximately 10 degrees. To remove the remaining parasitic light propagating within this cone, an interference filter of bandwidth 10 nm is placed in front of the hole. The laser source is a standard and safe laser diode with power 0.9 mW and wavelength 650 nm. An adjustable focusing lens is mounted in front of the laser to adjust the spot size on the product, typically to a few mm.

The camera has a 320x240 pixel sensor with a maximum frame rate of 30 images/s. The slowest micro-structure movement detectable by HORUS® is ultimately limited by mechanical stability. The minimum measurable speckle rate is around $10^{-5}$ Hz, corresponding to a correlation time $\tau_c$ of 100000 seconds, slightly more than 1 day. The maximum accessible speckle rate is limited by the camera speed itself. As explained below, the algorithm imposes the capture of a minimum of 4 images in order to calculate the SR: the maximum SR is thus limited to about 10Hz.

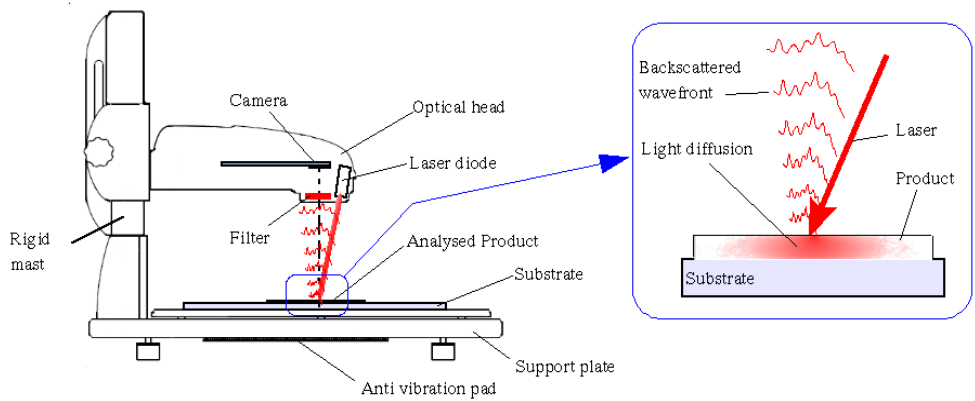

Fig. 1. The HORUS® setup.

## 3. The ASII method: inter-image distance, exponential approximation and adaptive algorithm

*3.1 Inter-image distance*

In order to minimize the computation load needed to monitor the drying process, one may wonder whether the evolution of the speckle pattern may be characterized more efficiently than by calculating the intensity correlation function $g_2$. A convenient way to address this question is to think at a speckle image as a point in a multi-dimensional space $S_{im}$. For any given image, each co-ordinate in the hyper-space $S_{im}$ corresponds to the intensity measured at a given pixel of the CMOS camera. The dimensionality of $S_{im}$ is thus the number of pixels of the camera, n=76800 for our 320x240 camera. The change of the speckle pattern between two images may then be quantified by introducing a vectorial distance function between the corresponding 2 points in the space $S_{im}$.

$$\text{electronic images} \longrightarrow \text{space } S_{im} \longrightarrow R^+$$
$$(\text{image1,image2}) \longmapsto (p_1,p_2) \text{ in } S_{im} \longmapsto \text{distance}$$

The distance is by definition a single non-negative number, which can be used to compare quantitatively pairs of speckle images taken at different times. If the two images are identical, the corresponding points in the hyper-space $S_{im}$ coincide and their distance is zero. As the two images become different, their distance will increase.

Various definitions for the distance function may be chosen; here we discuss briefly two possible choices that are computationally appealing. Let's note $I_1(x,y)$ the intensity of a pixel of co-ordinates (x,y) of the first image and $I_2(x,y)$ the corresponding intensity for the second image. We define a quadratic vectorial distance $d_2=\{\Sigma[I_2(x,y)-I_1(x,y)]^2\}^{1/2}$, as well as an alternative distance $d_1=\Sigma|I_2(x,y)-I_1(x,y)|$. In both cases, the sum is over all pixels. The inter-image distance has an asymptotic maximum value that we shall call $d_{max}$, corresponding to two fully uncorrelated speckle images. As we show below, both $d_1$ and $d_2$ are less demanding in terms of computational power than the traditional correlation function $g_2$. In the HORUS® apparatus we have chosen to implement $d_2$, which, as we will discuss below, is directly related to the more common $g_2$ function. Note that this quadratic inter-image distance represents the extension to multi-speckle processing of the "structure function" introduced by Schätzel for point-like detectors, which was shown to be less prone to detector noise than $g_2$ [19].

*3.2 Comparison of the computational load*

For the case of an ideal, noise-free detector with *n* pixels, computing $g_2$ requires *n* multiplications, while $d_2$ needs *n* differences and *n* squares, and $d_1$ needs *n* differences and *n* absolute values. Computing an addition is faster than a multiplication and easier to implement in a specialized electronics. Taking a square can be at least as fast as an addition thanks to a (small) look-up table [20]. Thus, the intensity correlation function $g_2$ is the most time consuming and the inter-image difference $d_1$ the less time consuming way to quantify the speckle pattern evolution.

The inter-image distance has also some additional advantages in the case of non-ideal detectors. A first typical imperfection is the fact that each pixel has its individual offset, or dark level. This offset must be corrected for when calculating $g_2$ [16], requiring two more additions per pixel, while it is automatically removed when using $d_2$ or $d_1$. Table 1 below recapitulates the computational cost of the various schemes discussed above, for the case of a camera having pixels with distinct offsets:

Table 1. Computational cost of various correlation tools, including camera offset calibration

| tool | additions or subtractions | squares | multiplications | Needs offset calibration |
| --- | --- | --- | --- | --- |

| tool | additions or subtractions | squares | multiplications | Needs offset calibration |
| --- | --- | --- | --- | --- |
| $g_2$ | $2n$ | 0 | $n$ | yes |
| $d_2$ | $n$ | $n$ | 0 | no |
| $d_1$ | $n$ | 0 | 0 | no |

It appears clearly that the distances $d_1$ or $d_2$ allow for a signal processing faster than that required for the traditional intensity auto-correlation.

Note that it is possible to establish a relation between $d_2$ and $g_2$ involving only on the mean intensity, $<I>$, and the mean squared intensity, $<I^2>$. By assuming that the probability distribution of the scattered intensity is the same for each pixel and using the classical relation between the variance and covariance of two random variables, one finds:

$$d_2 = [2n(<I^2> - g_2 <I>^2)]^{1/2} \qquad (1)$$

In particular, the maximum inter-image distance corresponding to two completely uncorrelated images (for which $g_2-1=0$) can be calculated using the previous relation, yielding $d_{2max} = [2n(<I^2> - <I>^2)]^{1/2}$. Therefore, $d_{2max}$ may be calculated directly from the probability distribution function (PDF) of the intensity obtained from one single speckle image, provided that the dark level offset is the same for all pixels.
If the speckle size is much larger than the pixel size, the PDF $P(I)$ of the scattered intensity is exponential: $P(I)=(1/<I>)\exp(-I/<I>)$ [21]. In this case $<I^2>=2<I>^2$ and the relationship between $g_2$ and $d_2$ simplifies to $g_2-1=1-(d_2/d_{2max})^2$, with $d_{2max}=(2n)^{1/2}<I>$.

*3.3 Estimation of the characteristic rate of change of the speckle field by an exponential approximation.*

In order to monitor a drying process, a series of short movies of the speckle pattern backscattered by the sample is acquired. Each movie is processed independently to extract the information related to the speckle dynamics. For each movie, the inter-image distance, $d_2(\tau)$, between the first image of the movie and all subsequent images, is calculated. Typical examples for three different drying fluids are shown in Fig. 2. For the sake of simplicity, it is desirable to extract only one single figure representative of the speckle dynamics for each movie. Experimentally, we have noticed that the shape of $d_2(\tau)$ is very often well described by an exponential growth to $d_{2max}$, the saturation level:

$$d_2(\tau) = d_2^0 + (d_{2max} - d_2^0)[1-\exp(\tau/\tau_c)] \qquad (2)$$

with $d_2^0$ the $d_2$ value for the first pair of images. Note that $d_2^0$ may be significantly larger than zero due to electronic noise and speckle dynamics faster than the minimum lag between images. It is therefore natural to fit the data to Eq. (2) and take the characteristic time $\tau_c$ for the typical decay time we are looking for. Finally, we define the "speckle rate" (SR) as the inverse of the correlation time $\tau_c$, SR=$1/\tau_c$, in order to provide a single number that decreases as the dynamics slows down.

Two practical issues are worth discussing. First, the asymptotic value of $d_2$ ($d_{2max}$) is required in Eq. (2). As discussed above, $d_{2max}$ could be calculated from the PDF of the speckle intensity for one single image. However, we find that more robust results are obtained by evaluating $d_{2max}$ using a pair of fully decorrelated speckle images, which we obtain from the first image and a software-rotated version of it. Second, we aim to extract the SR as rapidly as possible, in order to follow the drying kinetics.

Once $d_{2max}$ is known, this may be achieved by acquiring a shorter speckle movie, covering only the initial, linear growth of $d_2$. $\tau_c$ is then obtained by intersecting the initial linear growth with the asymptote $d_{2max}$. Practically, a linear adjustment of the $d_2$ data is performed and extrapolated until the horizontal line $d_2 = d_{2max}$ is crossed. This fitting procedure filters out most of the measurement noise. We have found that acquiring speckle images during $\tau_c/4$ yields satisfactory results. Therefore, the SR can be measured in a quarter of the speckle field characteristic time.

*3.4 Adaptive algorithm*

Ideally, each movie has a duration $T=\tau_c/4$ and contains as much images as possible. Since the dynamics is usually non-stationary, the value of $T$ must be permanently adapted before acquiring the next movie. We start by an initial guess $T_1 = 100$ ms, and calculate the first value of the SR, $SR_1$, as explained above. $SR_1$ is used as an input to calculate the duration of the following movie, which is set to $T_2 = 1/(4SR_1)$. This procedure is iterated during the whole measurement. Once the duration $T_i$ of the $i$-th has been set, the frame rate for the $i$-th movie is chosen as large as possible, within the limits of the camera (30 Hz) and with the additional constraint that each movie contains at most 255 images (for computational efficiency).

This algorithm acts as an efficient feed-back loop: if the first guess is not correct, the algorithm converges rapidly towards the correct one. If $\tau_c$ evolves during the drying process, the algorithm will adapt the acquisition and processing to the new one. Even if $\tau_c$ drops abruptly (e.g. if the speckle field "freezes" rapidly), the reduced time needed to evaluate the SR ($\tau_c/4$) allows a fast convergence to be achieved.

**4. Experimental tests**

*4.1 Examples of inter-image distance curves*

Figure 2 shows three typical examples of normalized inter-image distance curves, together with fits according to Eq. (2) . The first curve to the left ($\tau_c=$ 1s) was recorded during the drying of a white correction fluid. The intermediate one ($\tau_c=$ 5s) is from an acrylic white paint, while the slowest drying process (right curve, $\tau_c=$ 9 s) was measured during the drying of a red marker pen ink on metal. Note that in all cases the data are in excellent agreement with the exponentially saturating growth, Eq. (2).

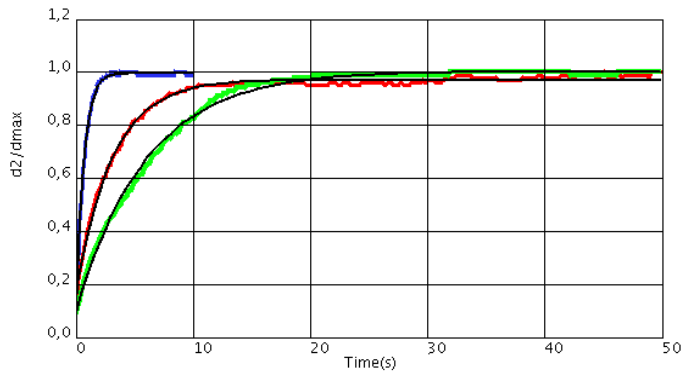

Fig 2. Normalized inter-image distance $d_2/d_{max}$ versus time for white corrector (blue line), acrylic paint (red line) and marker pen ink (green line). The black lines are fits according to Eq. (2).

*4.2 Examples of speckle rate kinetics curve*

We present in Fig. 3 three examples of drying experiments performed with HORUS®. The first two examples concern the film formation of a paint layer of 40 µm of thickness. The first one is an organic-solvent-based paint applied on a glass substrate (blue curve in Fig. 3). The SR *vs* time curve exhibits a first plateau at 2000 s, which corresponds to the end of volatile solvent evaporation, or "dry-to-touch" time in the paint industry language. The speckle rate reaches a final plateau at $10^{-4}$ Hz after 22000 s, when the paint is "dry-hard". The orange curve in Fig. 3 shows the drying process of a water-based latex paint applied to a metal substrate. Note the large fluctuations of the speckle rate around 200 s, likely due to the packing process of the colloidal particles contained in the paint during the film formation. This temporally heterogeneous dynamics is strongly reminiscent of that of jammed colloids [22]. The "dry-to-touch" plateau is reached at 800 s and the "dry-hard" plateau is reached after 5000 s. The third experiment shown in Fig. 3 (red curve) is the drying of a mark left by a commercial pen marker on a glassed paper. Collectively, the data shown in Fig. 3 illustrate the capability of HORUS® to follow both very slow or rapid drying processes and to capture even quite abrupt changes in the speckle rate evolution, as, e.g., the sharp drop of the SR around $t = 10$ s for the pen marker ink.

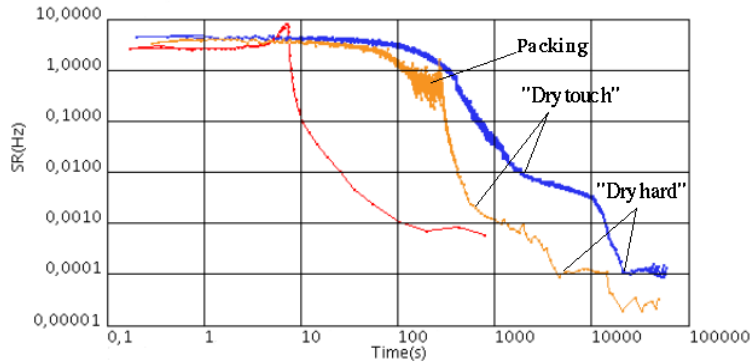

Fig. 3. Rate of change of the speckle pattern, SR, versus time in a double logarithmic graph. The fastest-drying curve (left, red) is for a commercial marker pen, the middle one is for a water-based paint (middle, orange) and the longest drying curve if for a organic solvent- based paint (right, blue).

*4.3 Penetration depth*

One question of practical importance is how deep in a drying film HORUS® can detect some motion: in other words, can this apparatus monitor the drying of the material below a superficial layer where a solid skin has been already formed. To address this issue, a simple experiment has been performed. A layer of perfectly dry paint with a known thickness and photon transport mean free path $l^*$ [4] is placed on a transparent glass substrate. Another layer of fresh paint is spread below the glass plate. The purpose of the experiment is to determine the maximum thickness, *L*, of dry paint beyond which HORUS® does not detect anymore the dynamics in the fresh paint. We find that *L* depends almost only on the photon transport mean free path $l^*$ of the dry paint, especially in the limit $L \gg l^*$. In particular, for $L/l^* > 30$ [23] the fresh layer is not detected anymore, indicating that its contribution to the inter-image distance becomes smaller than the measurement noise. The penetration depth *L* is then at least 30 $l^*$, which corresponds to 150 µm to 600 µm for a classical white paint

(the *l\** value for a paint goes from 5 to more than 20 μm).

*4.4 Influence of concentration, thickness and support*

In order to check the impact on the measurement of the film thickness and of the particle content, we have prepared two suspensions of Titanium dioxyde particles in castor oil (viscosity: 1 Pa.s at 21 °C): one with a volume fraction Φ=1 % (*l\**=300 μm) and one with Φ=0.1 % (*l\**=3000 μm). Note that since the boiling point of castor oil is around 300 ºC, the product is not drying at all during the experiment: the speckle rate SR is thus stationary. To avoid border effects, a large stripe of width 5 cm is spread on the substrate using a special tool. For the first suspension (Φ=1 %) spread on a metallic substrate, a layer of thickness 1000 μm gives a speckle rate of 7 Hz, 250 μm gives 3.2 Hz and 80 μm gives 1.5 Hz. For the second suspension (Φ=0.1 %) also spread on a metallic substrate, the layer of thickness 1000 μm gives a speckle rate of 0.5 Hz, 250 μm gives 0.4 Hz and 80 μm gives 0.02 Hz. Therefore, decreasing the layer thickness or particle volume fraction reduces the mean number of photon scattering events in the film [5] and hence decreases the speckle rate. This example shows that it is mandatory to control carefully the layer thickness when comparing absolute SR's measured for different products.

To check the effect of the substrate color or albedo on the measured SR, we have performed measurements for two different substrates: the black and the white part of a standard paper test chart used in paint industry. The albedo for the wavelength used is almost one for the white part and 0.125 for the black substrate. For the black substrate, the majority of the photons reaching the substrate are absorbed. By contrast, part of the photons reaching the white substrate diffuse in the (static) structure of the substrate and eventually are backscattered through the paint to the detector. The drying process of a layer of thickness 5 μm of a white acrylic paint (*l\** about 15 μm and *L/l\**=5) is measured for both substrates. Monte Carlo simulations show that the mean number of backscattered photons is about 3 times higher for the white substrate than for the black one. As a consequence, one expects a speckle rate about 3 times lower for the black substrate than for the white one. Indeed, the SR kinetics measured for the two substrates have the same shape, but the one corresponding to the black substrate is lower by a factor of about 3 with respect to that obtained for the white substrate, as shown in Fig. 4. Therefore, the substrate color or albedo does not affect the shape of the drying curve, an important feature in industrial applications.

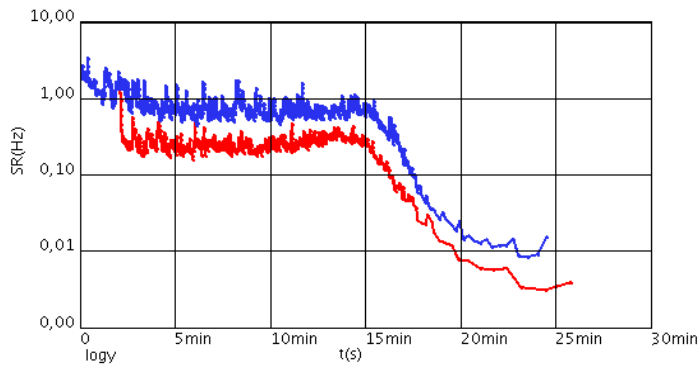

Fig. 4. Rate of change of the speckle pattern, SR, versus time in a semi logarithmic graph, for the drying of a 75μm layer of an acrylic paint. The blue curve (higher SR) corresponds to a white substrate and the red curve

(lower SR) corresponds to a black substrate. There is a ratio of approximately 3 between the absolute magnitudes of the 2 curves, but the shape of SR vs t is insensitive to the substrate color.

## 5. Conclusion

We have described and demonstrated an extension of multi-speckle diffusing wave spectroscopy adapted to characterize in a simple way the kinetics of micro-structure dynamics during drying processes. This method allows one to follow the evolution of a drying process in a very reactive way and with a high dynamic range. We call this technique ASII for "Adaptive Speckle Imaging Interferometry", a method based on the efficient quantification of the speckle pattern evolution (via the inter-image distance $d_2$), its characterization through a single number (the speckle rate SR) and an adaptive algorithm that continuously optimizes the acquisition and processing chain with respect to the product dynamics. These new methods are implemented in a novel commercial instrument, called HORUS®, specialized in the non-invasive analysis of the film formation and drying processes analysis.